\documentclass[twocolumn]{aastex63}
\usepackage{amssymb,amsmath}
\usepackage{graphicx}
\usepackage{xcolor,natbib}
\usepackage{multirow}
\usepackage[T1]{fontenc}
\usepackage{ae,aecompl}
\usepackage{newtxtext, newtxmath}
\usepackage{float}
\usepackage{subfigure}
\usepackage{longtable}
\usepackage{enumitem}
\usepackage{longtable,listings}
\usepackage[flushleft]{threeparttable}
\usepackage{parcolumns}
\def\msig{$M_{\rm BH}- \sigma$\ }

\def\obj{{\it Swift} J2058.4+0516}

\shorttitle{BH mass of {\it Swift} J2058.4+0516}
\shortauthors{ZHANG}

\begin{document}

\title{Central black hole mass in the distant tidal disruption event candidate 
of {\it Swift} J2058.4+0516}

\correspondingauthor{XueGuang Zhang}%
\email{xgzhang@njnu.edu.cn}

\author{XueGuang Zhang$^{*}$}
\affiliation{School of Physics and technology, Nanjing Normal University, No. 1,
        Wenyuan Road, Nanjing, 210023, P. R. China}

\begin{abstract} 
	In the manuscript, central black hole (BH) mass is estimated in the distant TDE (tidal 
disruption event) of {\it Swift} J2058.4+0516 as the second candidate of relativistic jet birth 
related to TDE. {\it Swift} J2058.4+0516 have quite different BH masses estimated through 
different indirect methods in the literature. Therefore, it is necessary and interesting to 
determine the central BH mass in {\it Swift} J2058.4+0516 by one another independent method. 
Here, based on the theoretical TDE model applied to describe the long-term time-dependent X-ray 
variabilities of {\it Swift} J2058.4+0516, the central BH mass can be well determined to be 
around $1.05_{-0.29}^{+0.39}\times10^5{\rm M_\odot}$, after kind considerations of 
suggested intrinsic beaming effects from such relativistic jet tightly related to TDE. 
Moreover, the {\it Swift} J2058.4+0516 is an unique object in the space of BH masses versus 
energy transfer efficiencies of the reported TDE candidates, providing interesting clues to 
detect and/or anticipate candidates of relativistic TDE to make the birth of relativistic jets. 
\end{abstract}


\section{Introduction}
 
	\obj\ (= {\it Swift} J20:58:19.90+05:13:32) has been reported as a Tidal Disruption Event 
(TDE) candidate in \citet{ck12} by the long-term X-ray variabilities discovered by 
the Burst Alert Telescope \citep{bb05} on the Swift satellite \citep{gc04}, and then by the 
long-term X-ray variabilities discussed and reported in \citet{pc15} and in \citet{wh20}, and 
also reported as the second candidate of relativistic TDE to make the birth of relativistic jets 
similar as the results in {\it Swift} J1644+57\ well discussed in \citet{bur11, zb11, bz12, mb16, 
eb18}. Moreover, as the shown results in Fig. 1\ in \citet{pc15}, there are similar time-dependent 
optical and X-ray variability properties. Therefore, significant correlation between X-ray and 
optical light curves can be confirmed in {\it Swift} J2058.4+0516, providing further evidence to 
support that it is also an efficient way to detect the TDE by X-ray properties in {\it Swift} 
J2058.4+0516, although most TDEs are commonly detected and reported by optical variability properties. 
Among the reported more than 80 TDE candidates \citep{ko99, ko04, ve11, gs12, md15, ht16, lz17, 
wz17, wy18, gr19, vv19, we19, an20, hi21} (see detail in \url{https://tde.space}) with a star tidally 
disrupted by central massive black hole (BH) leading to time-dependent structure evolutions of 
accreting flows around the central BH, \obj~ is the target in the manuscript, in order to find 
more clues on properties of central BH mass of \obj, mainly because of the inconsistent BH masses 
of \obj\ reported in the literature.

	In he literature, there are quite different reported BH masses in \obj. \citet{ck12} have 
discussed that the central BH mass of \obj\ should be smaller than $10^8{\rm M_\odot}$ after 
considerations of the shortest X-ray variability time scale in \obj\ and considerations of the 
BH fundamental plane discussed in \citet{gul09}. \citet{pc15} have reported the central BH mass 
in \obj\ to be in the range from $10^4{\rm M_\odot}$ to $10^6{\rm M_\odot}$, after analyzing 
properties of the sudden decline with proposed transition from super-Eddington to sub-Eddington 
accretion in \obj. Meanwhile, if a white dwarf rather than a main-sequence star was tidally 
disrupted in central region of \obj\ as discussed in \citet{kp11} for {\it Swift} J1644+57, the 
central BH mass could be smaller than $10^5{\rm M_\odot}$ in \obj. More recently, after 
considerations of X-ray spectral properties, \citet{stv17} have shown that the central BH mass 
of \obj\ should be larger than $2\times10^7{\rm M_\odot}$ with redshift $z=1.1853$ accepted, 
through the scaling technique \citep{st09, ts17} applied with the Galactic BHs, 
GRO J1655-40, GX 339-4, Cyg X-1 and 4U 1543-47 as reference sources. It is clear that different 
techniques/methods with different considerations lead to very different central BH masses of 
\obj. Once, there was a more robust technique applied to determine the accurate central BH mass 
of \obj, central engine could be clearer enough in \obj. Therefore, it is important and interesting 
to determine the central BH mass in the TDE of \obj\ through one another independent method.

	Commonly, there are some convenient methods/techniques applied to measure/estimate BH 
masses of AGN and/or quiescent galaxies, such as the well-known \msig relation \citep{fm00, ge00, 
kh13, sg15} and the well-applied Virialization method \citep{pe04, sr10, rh11}. More recently, 
TDEs properties can be well applied to determine central BH masses, such as the results shown 
and discussed in \citet{gm14, mg19, zl21}, etc. Due to weak host galaxy spectroscopic absorption 
features and lack of broad emission lines, neither the \msig relation nor the Virialization method 
can be applied to measure/estimate the central BH mass in \obj. However, to accept a TDE with a 
central main-sequence star tidally disrupted to describe the observed long-term X-ray variabilities 
in \obj, the central BH mass of \obj~ can be well determined, which is the main objective of 
the manuscript. Section 2 presents the methods and model to describe X-ray variabilities 
of \obj~ by the theoretical TDE model. Section 3 gives our main results and necessary discussions. 
Section 4 gives our summaries and conclusion. In the manuscript, we have adopted the 
cosmological parameters $H_{0}=70{\rm km\cdot s}^{-1}{\rm Mpc}^{-1}$, $\Omega_{\Lambda}=0.7$ and 
$\Omega_{\rm m}=0.3$.

\begin{figure*}
\centering\includegraphics[width = 18cm,height=6cm]{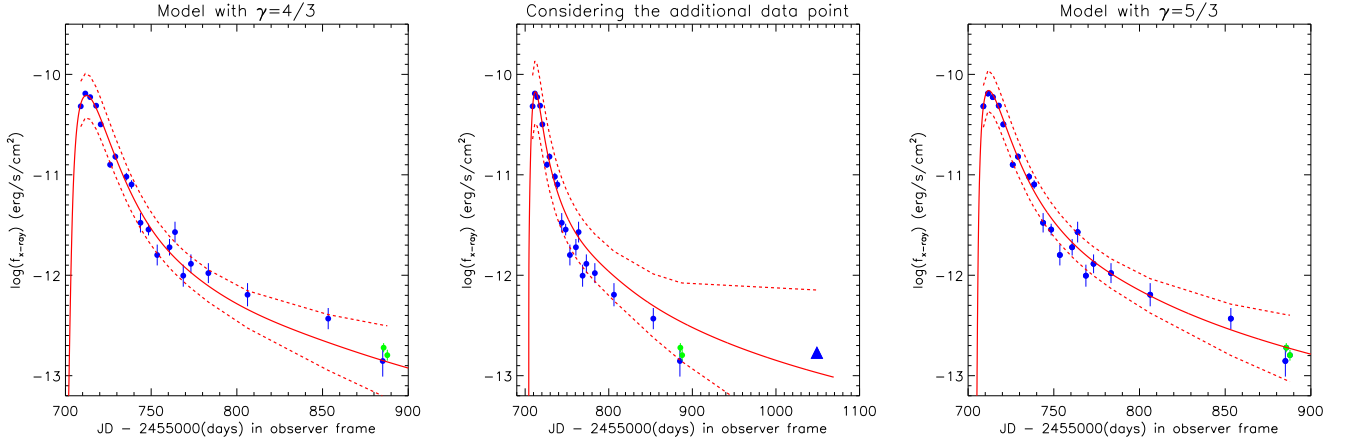}
\caption{Left panel shows the X-ray light curve in the bandpass of 0.3-10keV of \obj~ 
with JD from 2455708.915 to 2455887.787, and the best descriptions (shown in solid red line) and 
the corresponding 99\% confidence bands (shown in dashed red lines) determined by the theoretical 
TDE model with $\gamma=4/3$. Middle panel shows the light curve including the additional data point 
shown as solid blue triangle with JD=2456049.048, and the best descriptions (shown in solid red 
line) and the corresponding 99\% confidence bands (shown in dashed red lines) determined by the 
theoretical TDE model with $\gamma=4/3$. Right panel shows the light curve with JD from 2455708.915 
to 2455887.787, and the best descriptions (shown in solid red line) and the corresponding 99\% 
confidence bands (shown in dashed red lines) determined by the theoretical TDE model with $\gamma=5/3$. 
In each panel, solid circles plus error bars in blue show the 20 data points acquired with the X-Ray 
Telescope on the Swift satellite, and solid circles plus error bars in green show the two data points 
acquired with the European Photon Imaging Camera on the XMM-Newton Observatory. In middle panel, 
the solid blue triangle plus error bars show the additional data point with JD=2456049.048 acquired 
with the European Photon Imaging Camera on the XMM-Newton Observatory.
}
\label{tde}
\end{figure*}
	
\begin{figure*}
\centering\includegraphics[width = 18cm,height=11cm]{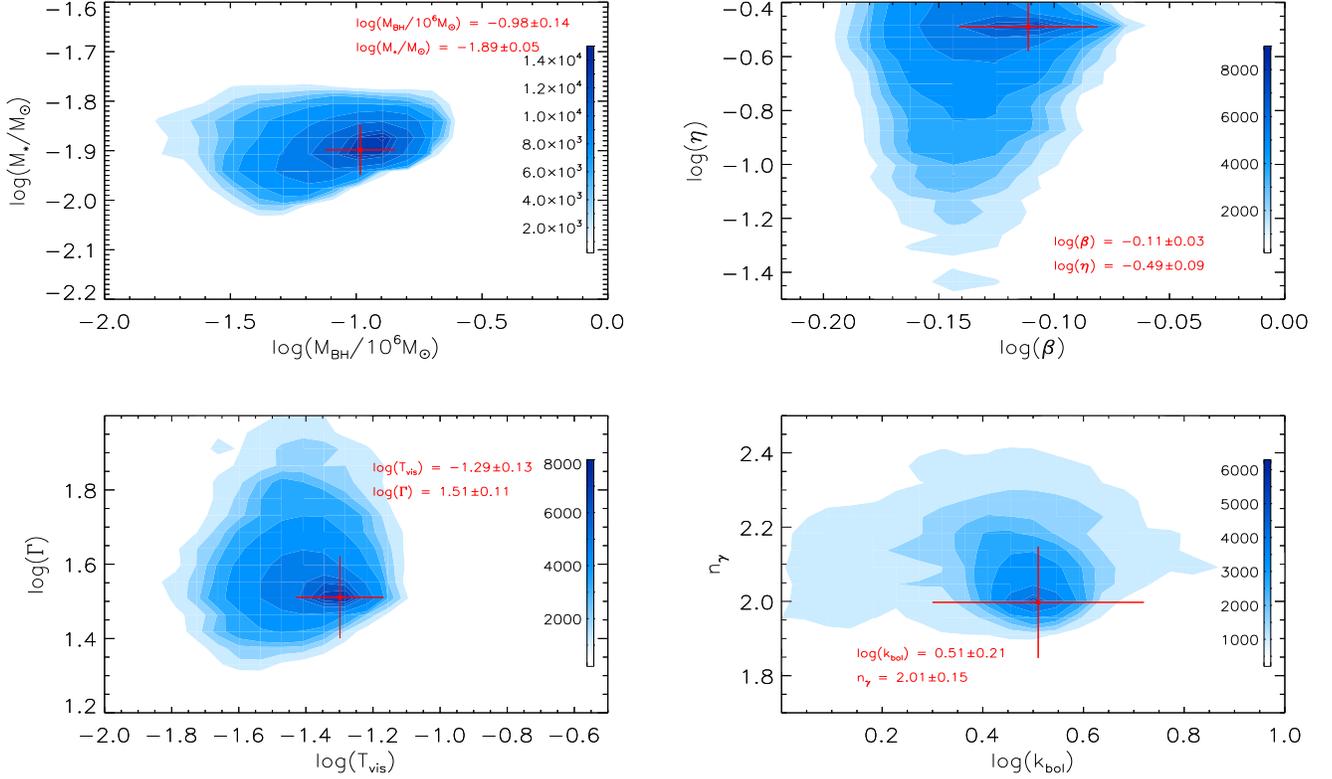}
\caption{The MCMC determined two-dimensional projections of the posterior distributions in contour 
of the eight model parameters of $\log(M_{\rm BH})$, $\log(M_{\star})$, $\log(\beta)$, $\log(\eta)$, 
$\log(T_{vis})$, $\log(k_{bol})$, $\log(\Gamma)$ and $n_\gamma$ through application of the theoretical 
TDE model with $\gamma=4/3$. In each panel, solid circle plus error bars in red show the accepted 
values and uncertainties of the model parameters which are also marked by red characters in each panel. 
}
\label{par}
\end{figure*}

\begin{figure*}
\centering\includegraphics[width = 16cm,height=10cm]{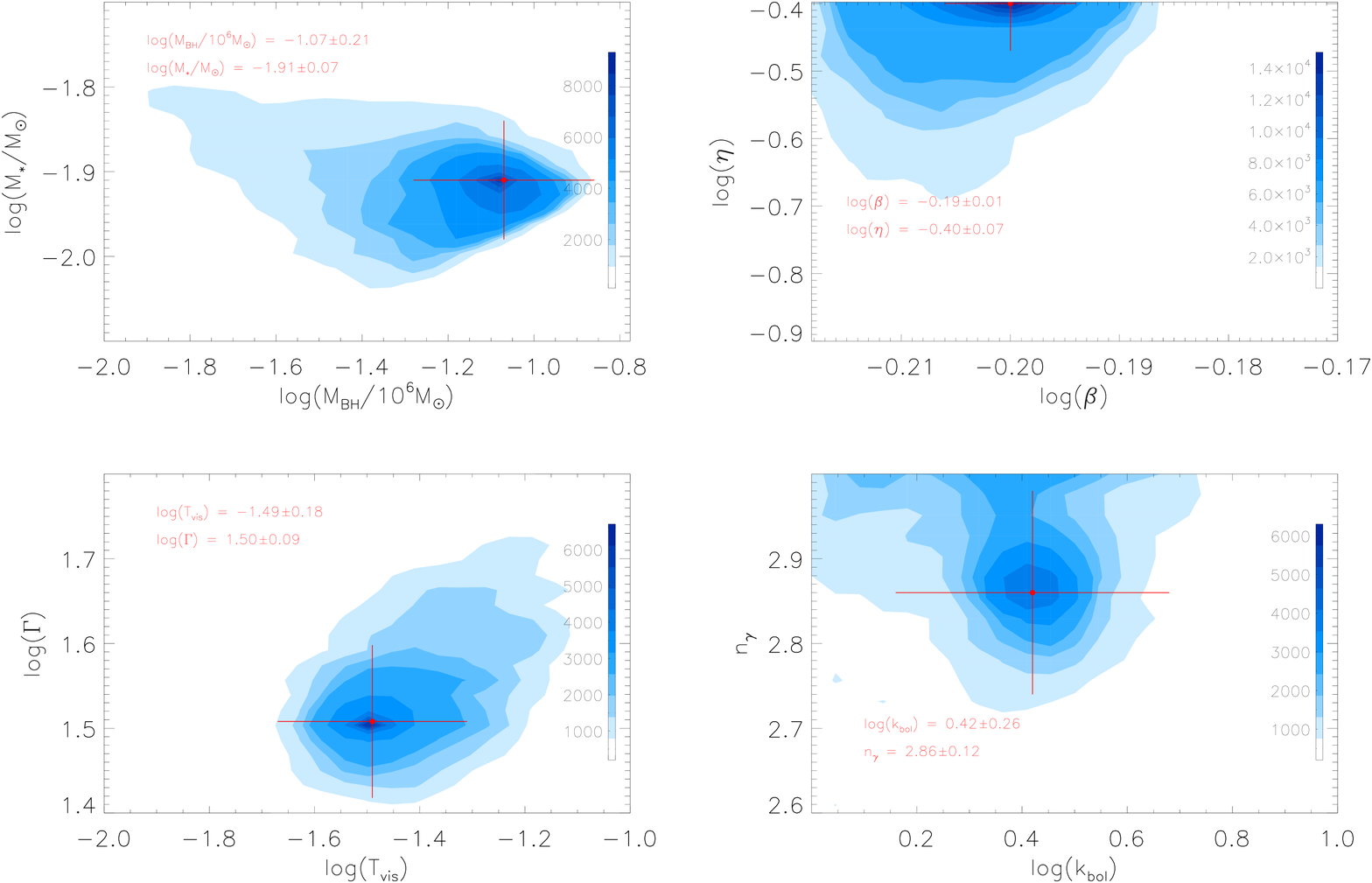}
\caption{Similar as Fig.~\ref{par}, but for the light curve including the additional data 
point with JD = 2456049.048 described by the theoretical TDE model with $\gamma=4/3$ shown in middle 
panel of Fig.~\ref{tde}.}
\label{p3ar}
\end{figure*}

\begin{figure*}
\centering\includegraphics[width = 16cm,height=10cm]{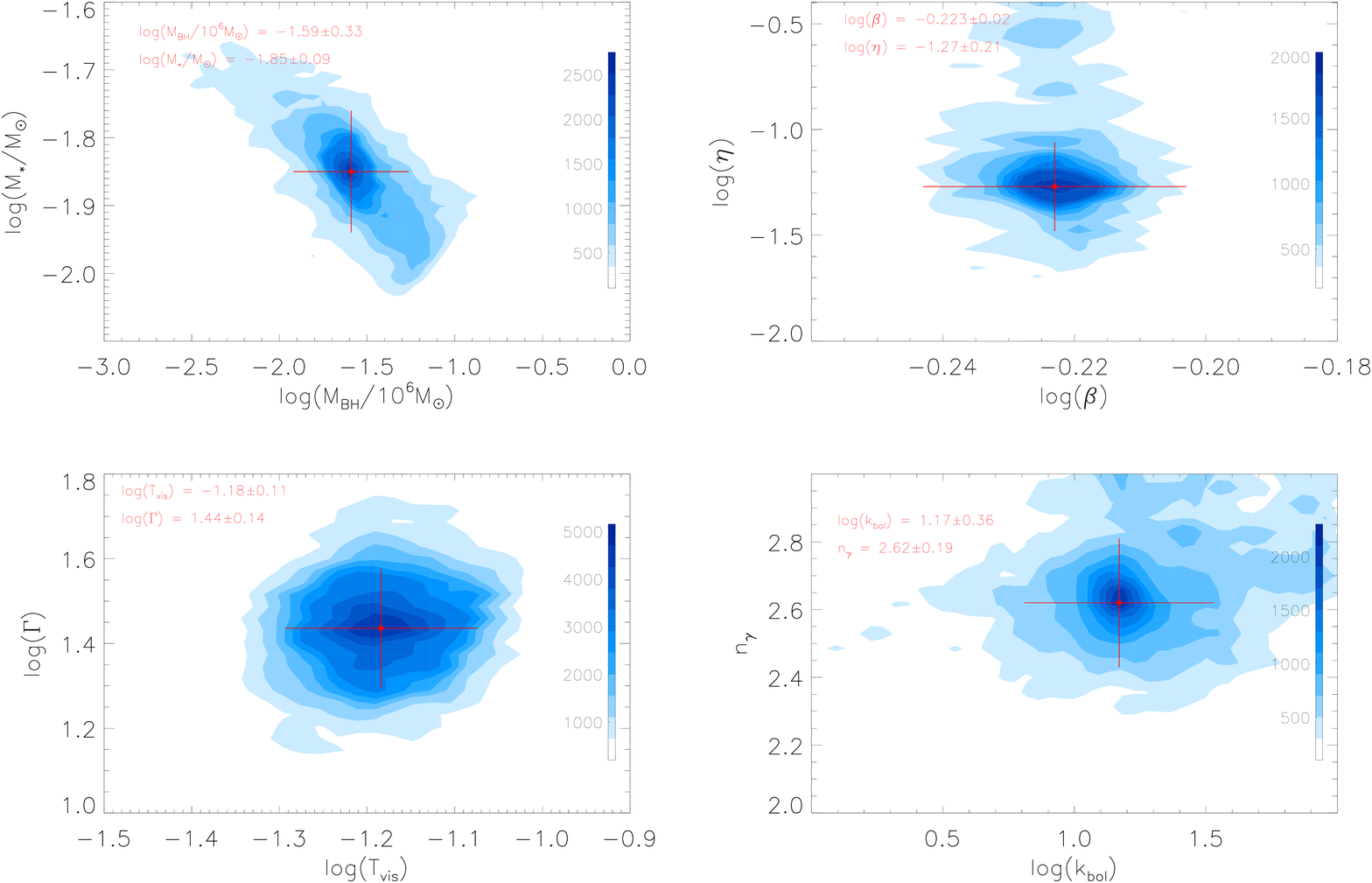}
\caption{Similar as Fig.~\ref{par}, but for the results shown in right panel of Fig.~\ref{tde} 
through application of the theoretical TDE model with $\gamma=5/3$.}
\label{p2ar}
\end{figure*}

\section{Methods and Model}


	In order to well describe the X-ray variabilities of \obj~ by the theoretical TDE  model, 
the following five steps are applied.

	First, standard templates of viscous-delayed accretion rates in TDEs are created. Based on 
the discussed $dM/dE$ provided in \citep{gm14, gn18, mg19}, templates of fallback material rate 
\begin{equation}
\dot{M}_{fbt}=dM/dE~\times~dE/dt
\end{equation}
can be created for the standard cases with central BH mass $M_{\rm BH}=10^6{\rm M_\odot}$ and 
disrupted main-sequence star of $M_{*}=1{\rm M_\odot}$ and with a grid of the listed impact parameters 
$\beta_{t}$ in \citet{gr13} with $dE/dT$ calculated by 
\begin{equation}
dE/dt~=~\frac{(2~\pi~G~M_{\rm BH})^{2/3}}{3~t^{5/3}}
\end{equation}.
Then, considering the viscous delay as discussed in \citet{gr13, mg19} by the viscous timescale 
$T_{vis}$, templates of viscous-delayed accretion rates $\dot{M}_{at}$ can be determined by
\begin{equation}
\dot{M}_{at}~=~\frac{exp(-t/T_{vis})}{T_{vis}}\int_{0}^{t}exp(t'/T_{vis})\dot{M}_{fbt}dt'
\end{equation}.
And a grid of 31 $\log(T_{vis,~t}/{\rm years})$ range from -3 to 0 are applied to create templates 
$\dot{M}_{at}$ for each impact parameter. The final templates of $\dot{M}_{at}$ include 736 (640) 
time-dependent viscous-delayed accretion rates for 31 different $T_{vis}$ of each 23 (20) impact 
parameters for the main-sequence star with polytropic index $\gamma$ of 4/3 (5/3).

	Second, for a TDE with model parameters of $\beta$ and $T_{vis}$ different from the list 
values in $\beta_{t}$ and in $T_{vis,~t}$, the actual viscous-delayed accretion rates $\dot{M}_{a}$ 
are created by the following two linear interpolations. Assuming that $\beta_1$, 
$\beta_2$ in the $\beta_{t}$ are the two values nearer to the input $\beta$, and that $T_{vis1}$, 
$T_{vis2}$ in the $T_{vis,~t}$ are the two values nearer to the input $T_{vis}$, the first 
linear interpolation is applied to find the viscous-delayed accretion rates with input 
$T_{vis}$ but with $\beta=\beta_1$ 
and $\beta=\beta_2$ by
\begin{equation}
\begin{split}
&\dot{M}_{a}(T_{vis},~\beta_{1})=\dot{M}_{at}(T_{vis1},~\beta_1) + \\
	&\ \ \ \frac{T_{vis}-T_{vis1}}{T_{vis2}-T_{vis1}}(\dot{M}_{at}(T_{vis2},~\beta_1)
	- \dot{M}_{at}(T_{vis1}, \beta_1))\\
&\dot{M}_{a}(T_{vis},~\beta_2)=\dot{M}_{at}(T_{vis1},~\beta_2) + \\
	&\ \ \ \frac{T_{vis}-T_{vis1}}{T_{vis2}-T_{vis1}}(\dot{M}_{at}(T_{vis2},~\beta_2)
	- \dot{M}_{at}(T_{vis1},~\beta_2))
\end{split}
\end{equation}.
Then, the second linear interpolation is applied to find the viscous-delayed accretion 
rates with input $T_{vis}$ and with input $\beta$ by
\begin{equation}
\begin{split}
&\dot{M}_{a}(T_{vis},~\beta)=\dot{M}_{a}(T_{vis},~\beta_1) + \\
	&\ \ \ \frac{\beta-\beta_1}{\beta_2-\beta_1}(\dot{M}_{a}(T_{vis},~\beta_2)
	- \dot{M}_{a}(T_{vis},~\beta_1))
\end{split}
\end{equation}.

	Third, for a TDE with input model parameters of $M_{\rm BH}$ and $M_{*}$ different from 
$10^6{\rm M_\odot}$ and $1{\rm M_\odot}$, the actual viscous-delayed accretion rates $\dot{M}$ and 
the corresponding time information in observer frame are created by the following scaling relations 
as shown in \citet{gm14, mg19},
\begin{equation}
\begin{split}
&\dot{M} = M_{\rm BH,~6}^{-0.5}~\times~M_{\star}^2~\times~
	R_{\star}^{-1.5}~\times~\dot{M}_{a}(T_{vis},~\beta) \\
&t_m = (1+z)\times M_{\rm BH,~6}^{0.5}~\times~M_{\star}^{-1}\times
	R_{\star}^{1.5}~\times~t_{a}(T_{vis},~\beta)
\end{split}
\end{equation},
where $M_{\rm BH,~6}$, $M_{\star}$, $R_{\star}$ and $z$ represent central BH mass in unit of 
${\rm 10^6M_\odot}$, stellar mass in unit of ${\rm M_\odot}$, stellar radius in unit of 
${\rm R_{\odot}}$ and redshift of host galaxy of a TDE, respectively. And the mass-radius relation 
well discussed in \citet{tp96} is accepted for main-sequence stars.

	Fourth, based on the calculated time-dependent templates of accretion rate $\dot{M}(t)$, the 
modeled time-dependent bolometric luminosity $L_{\rm bol,~t}$ can be well determined. Meanwhile, 
accepted the strong correlation between X-ray luminosity and bolometric luminosity, the 
TDE model expected X-ray band flux $f_{\rm X-ray,~m}$ can be simply calculated as
\begin{equation}
\begin{split}
&L_{\rm bol,~t}~=~\eta~\times~\dot{M}(t)\times c^2 \\
&f_{\rm X-ray,~m}~=~\frac{1}{k_{\rm bol}}~\times~\frac{L_{\rm bol,~t}}{4~\pi~D_{\rm obj}^2}
\end{split}
\end{equation}, 
where $c$, $\eta$, $k_{\rm bol}$ and $D_{\rm obj}$ represent the light speed, energy transfer 
efficiency around BH, the corresponding expected bolometric correction factor 
$k_{\rm bol}~=~\frac{L_{\rm bol}}{L_{\rm X-ray}}$ (where $L_{\rm bol}$ and $L_{\rm X-ray}$ as bolometric 
luminosity and X-ray band luminosity), and the distance of the \obj which can be well estimated by 
the redshift, respectively.

	Fifth, as results discussed on birth of relativistic jets related to TDEs \citep{bur11, zb11, 
bz12}, a free model parameter $\Gamma$, the relativistic Lorentz factor, has been considered to 
correct the beaming effects in TDE model simulated time duration $t$ and flux density $f$ by
\begin{equation}
\begin{split}
&t_{re}~=~t_{m}~\times~\Gamma \\
&f_{re}~=~f_{m}~\times~\Gamma^{n_{\gamma}}
\end{split}
\end{equation},
where "t" and "f" mean the time information and corresponding X-ray band fluxes, subscripts of "m" 
and "re" represent values from the common theoretical TDE model and the values after considering the 
relativistic beaming effects, respectively. Here, $n_{\gamma}$ is the well discussed parameter in 
\citet{cl07}. Moreover, as discussed results in \citet{ck12} for \obj~ and discussed results in 
\citet{bur11} for the first relativistic TDE in {\it Swift} J1644+57 to make the birth of relativistic 
jets, the reasonable $\Gamma$ could be larger than 2.1 and/or around 10-20.

	Finally, the theoretical TDE model expected time dependent X-ray light curve can be described 
by the following eight model parameters, the central BH mass $M_{\rm BH}$, the stellar mass $M_\star$ 
(corresponding stellar radius $R_\star$ tied by the mass-radius relation), the energy transfer efficiency 
$\eta$, the impact parameter $\beta$, the viscous timescale $T_{vis}$, the bolometric correction 
$k_{\rm bol}$, the Lorentz factor $\Gamma$ and the parameter $n_\gamma$. Meanwhile, the parameter of 
redshift of \obj~ has been determined and accepted as $z=1.1853$, through the spectroscopic features 
of Fe~{\sc ii} and Mg~{\sc i} shown in \citet{ck12}.

\section{Main Results and Necessary Discussions}

\subsection{Fitting the X-ray light curve}

	The reported X-ray light curve in the bandpass of 0.3-10keV of \obj~ in \citet{pc15} is shown 
in left panel of Fig.~\ref{tde} in observer frame. Similar long-term X-ray variabilities 
can also be found in the website of \url{https://tde.space} and in \citet{ck12}. It is clear that the 
smoothly declined trend was probably similar as TDE model expected time-dependent variabilities, as the 
results shown in \citet{ck12}: "the time-dependent X-ray variabilities can be described as 
$L_{X-ray}~\propto~t^{-2.2}$, with slope of $-2.2$ slightly different from the standard value of $-5/3$". 
As discussed in \citet{gr13}, only the extremely ideal cases for TDE with constant binding energy of the 
fallback materials to the BH $dM/dE~=~constant$ ($M$ as debris mass and $E$ the specific binding energy), 
the power law of accretion rates (luminosity) on time can be well expected by $t^{-5/3}$. However, for 
TDEs with $dM/dE$ depending on $E$, the expected power law $t^{\alpha}$ could have $\alpha$ different 
from $-5/3$, such as the detailed discussions in \citet{gr13}, power law with $\alpha~\sim~-2.2$ steeper 
than standard $-5/3$ can be well accepted in TDEs.

	There are 22 data points shown as solid circles in left panel of Fig.~\ref{tde}. Detailed 
descriptions on the 22 data points can be found in \citet{pc15}, simple descriptions on the 22 data 
points are described as follows. The 22 data points were acquired with two different instruments, 20 
data points (shown as solid blue circles) with JD from 2455708.915 to 2455885.11\ are from the X-Ray 
Telescope \citep{bu05} on the Swift satellite \citep{gc04}, and the other two data points (shown as 
solid green circles) are from the European Photon Imaging Camera \citep{st01, tu01} on the XMM-Newton 
Observatory \citep{ja01}. Moreover, there is one additional data point marked as solid blue triangle 
with JD=2456049.048 shown in middle panel of Fig.~\ref{tde}, which was acquired with the European 
Photon Imaging Camera on the XMM-Newton Observatory and will be discussed individually in the following 
paragraphs. Now, it is interesting to show that the theoretical TDE model can be applied to well 
describe the X-ray light curve of \obj~ by the procedure described in Section 2.

	Then, based on the theoretical TDE model described in Section 2, the observed X-ray light 
curve in the bandpass of 0.3-10keV of \obj~ in observer frame shown in left panel of 
Fig.~\ref{tde} can be well described by the TDE model with the tidally disrupted main-sequence star 
with polytropic index $\gamma=4/3$, through the well-known maximum likelihood method combining with 
the Markov Chain Monte Carlo (MCMC) technique \citep{fh13}. And the corresponding 99\% 
confidence bands are also shown in the left panel to the best descriptions to the X-ray light curve. 
When the theoretical TDE model is applied, there is only one criterion to limit the model parameters. 
For an available TDE with model parameters, the determined tidal disruption radius 
$R_{\rm TDE}$,
\begin{equation}
\frac{R_{\rm TDE}}{R_{\rm s}} = 5.06\times(M_\star)^{-1/3}(\frac{M_{\rm BH,~6}}
	{10})^{-2/3}\times R_\star > 1
\end{equation},
is larger than event horizon of central BH ($R_{\rm s}=2GM_{\rm BH}/c^2$).

	And the model parameters with accepted prior uniform distributions as follows, when the 
fitting procedure is running. The BH mass $\log(M_{BH}/10^6{\rm M_\odot})$ has a prior uniform 
distribution from -3 to 3 and has 0 as the starting value. The stellar mass $\log(M_{\star}/M_\odot)$ 
has a prior uniform distribution from -2 to 1.78 (about $60{\rm M_\odot}$) and has 0 as the starting 
value. The Lorentz factor $\log(\Gamma)$ has a prior uniform distribution from 0 to 1.7 and has 1 
as the starting value. The impact parameter of $\log(\beta)$ has a prior uniform distribution from 
$\log(0.6)$ to $\log(4)$ for main sequence stars with polytropic index of $4/3$ and has 0 as the 
starting value. The parameter of $\log(k_{\rm bol})$ has a prior uniform distribution from $\log(2)$ 
to $\log(100)$ and has 2 as the starting value. The parameter of $\log(T_{vis})$ has a prior 
uniform distribution from -3 to 0 and has -1 as the starting value. The parameter of $\log(\eta)$ 
has a prior uniform distribution from $\log(0.005)$ to $\log(0.4)$ and has $\log(0.15)$ as the 
starting value. And the parameter of $n_\gamma$ has a prior uniform distribution from 1 to 5 and 
has 2 as the starting value. Moreover, not similar as the different mass limits in 
\citet{gr13, mg19} for main-sequences stars with different polytropic indices, the same mass limits 
are accepted for main sequence stars with different polytropic indices by the following main 
consideration. \citet{gr13, gm14, mg19} have created the templates of $dM/dE$ for the cases with 
$\gamma=5/3$ after considerations of a main sequence star with stellar mass about one solar mass, 
indicating main sequence stars with $\gamma=5/3$ but with masses larger than 0.3 solar masses and 
smaller than 22 solar masses could be theoretically accepted. Therefore, similar mass limits are 
accepted for main sequence stars with different polytropic indices in the manuscript.

\begin{figure}
\centering\includegraphics[width = 8cm,height=5.5cm]{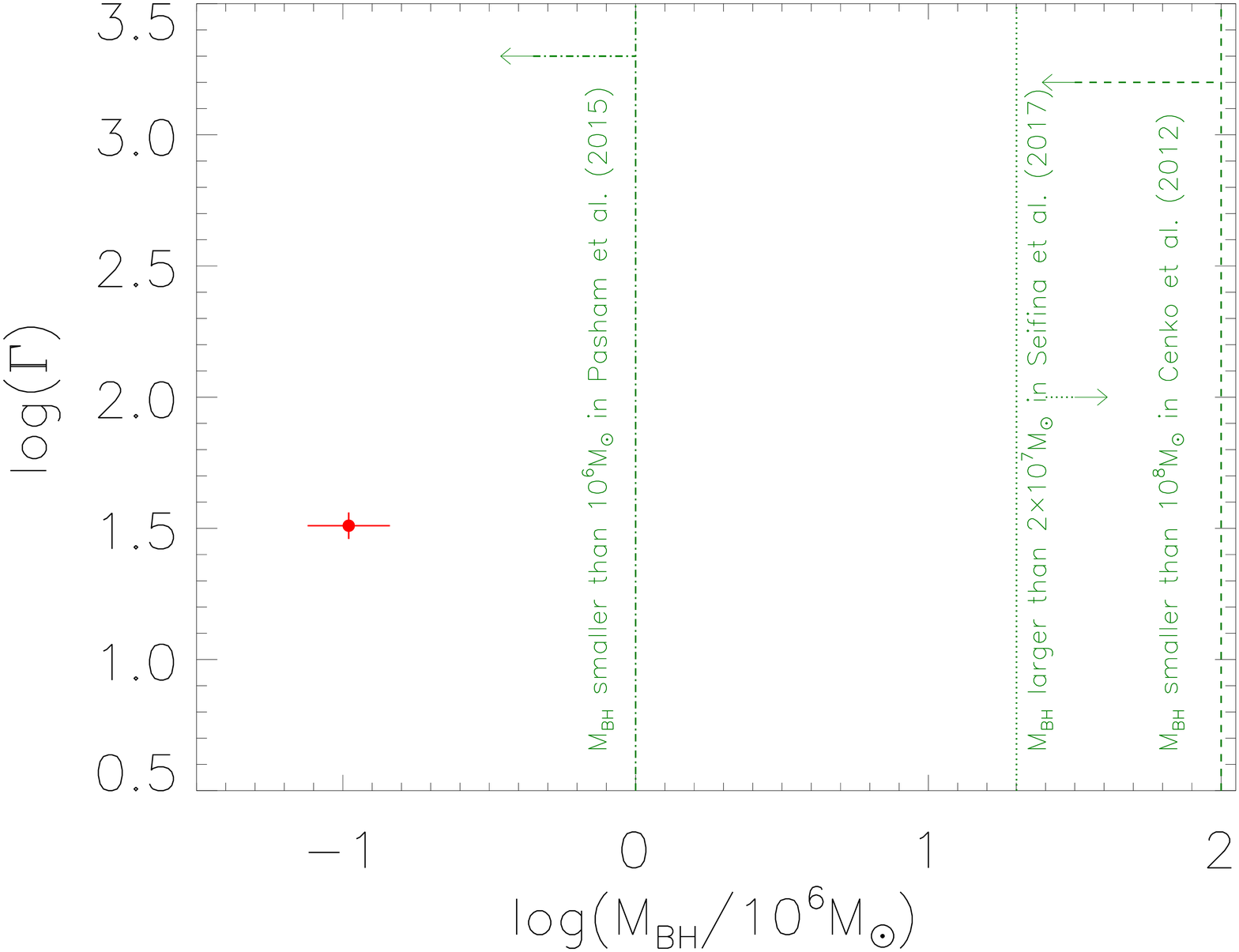}
\caption{Properties of BH mass of \obj~ determined by the theoretical TDE model and reported in the 
literature. Solid circle plus error bars in red show the results determined by the theoretical TDE 
model in the manuscript. The vertical dotted line in dark green shows the lower limit of central BH 
mass of \obj~ discussed in \citet{stv17}: BH mass larger than $2~\times~10^7{\rm M_\odot}$. The 
vertical dashed line in dark green shows the upper limit of central BH mass of \obj~ discussed in 
\citet{ck12}: BH mass smaller than $10^8{\rm M_\odot}$. The vertical dot-dashed line in dark green 
shows the upper limit of central BH mass of \obj~ discussed in \citet{pc15}: BH mass between 
$10^4{\rm M_\odot}$ to $10^6{\rm M_\odot}$.}
\label{all}
\end{figure}

\begin{figure}
\centering\includegraphics[width = 8cm,height=5.5cm]{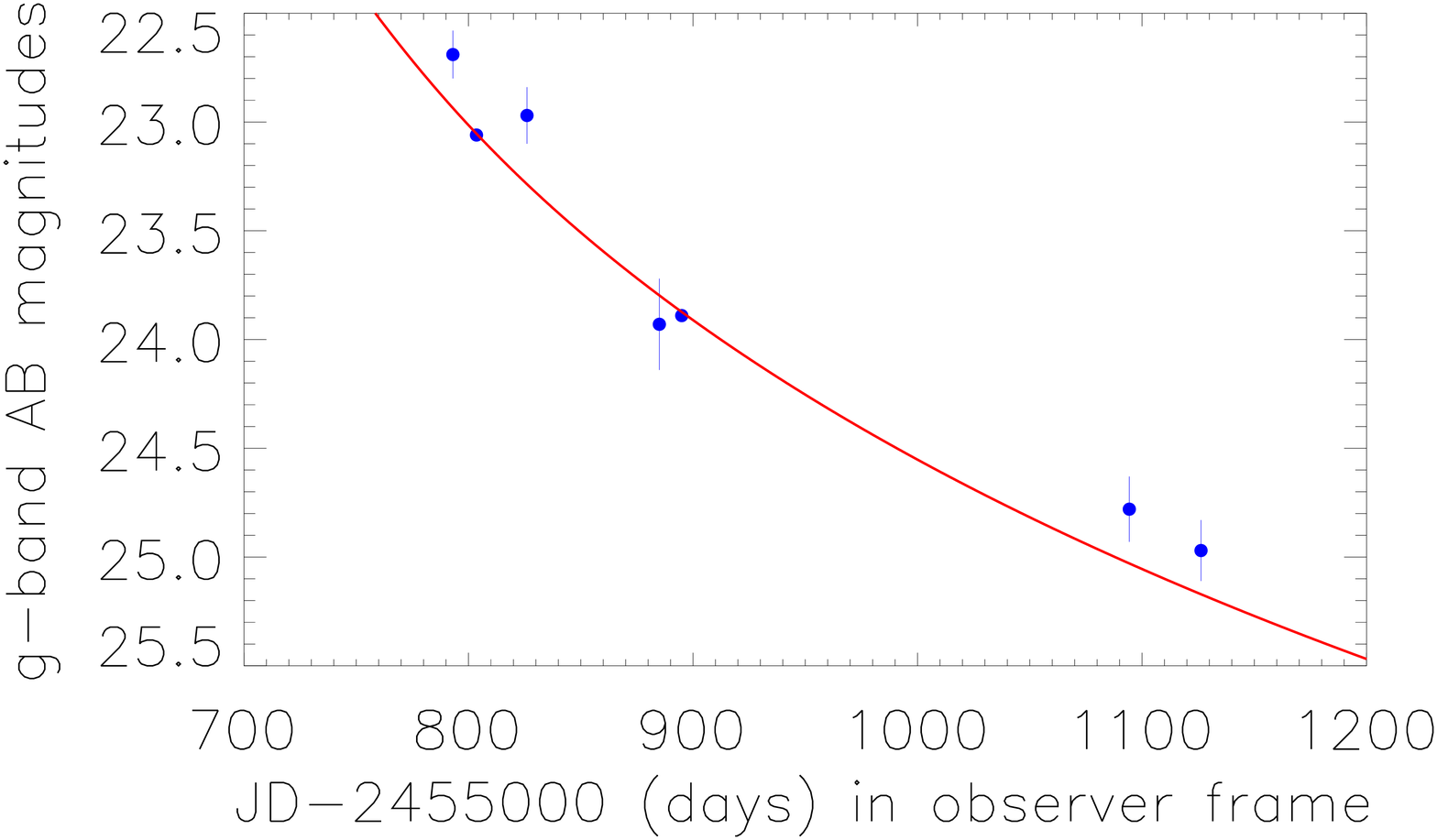}
\caption{The g-band optical light curve (solid circles plus error bars in blue) of \obj~ 
in observer frame, and the descriptions (solid red line) to the optical light curve by the shifted 
and weakened best descriptions to the X-ray light curve.}
\label{opt}
\end{figure}

\begin{figure}
\centering\includegraphics[width = 8cm,height=5.5cm]{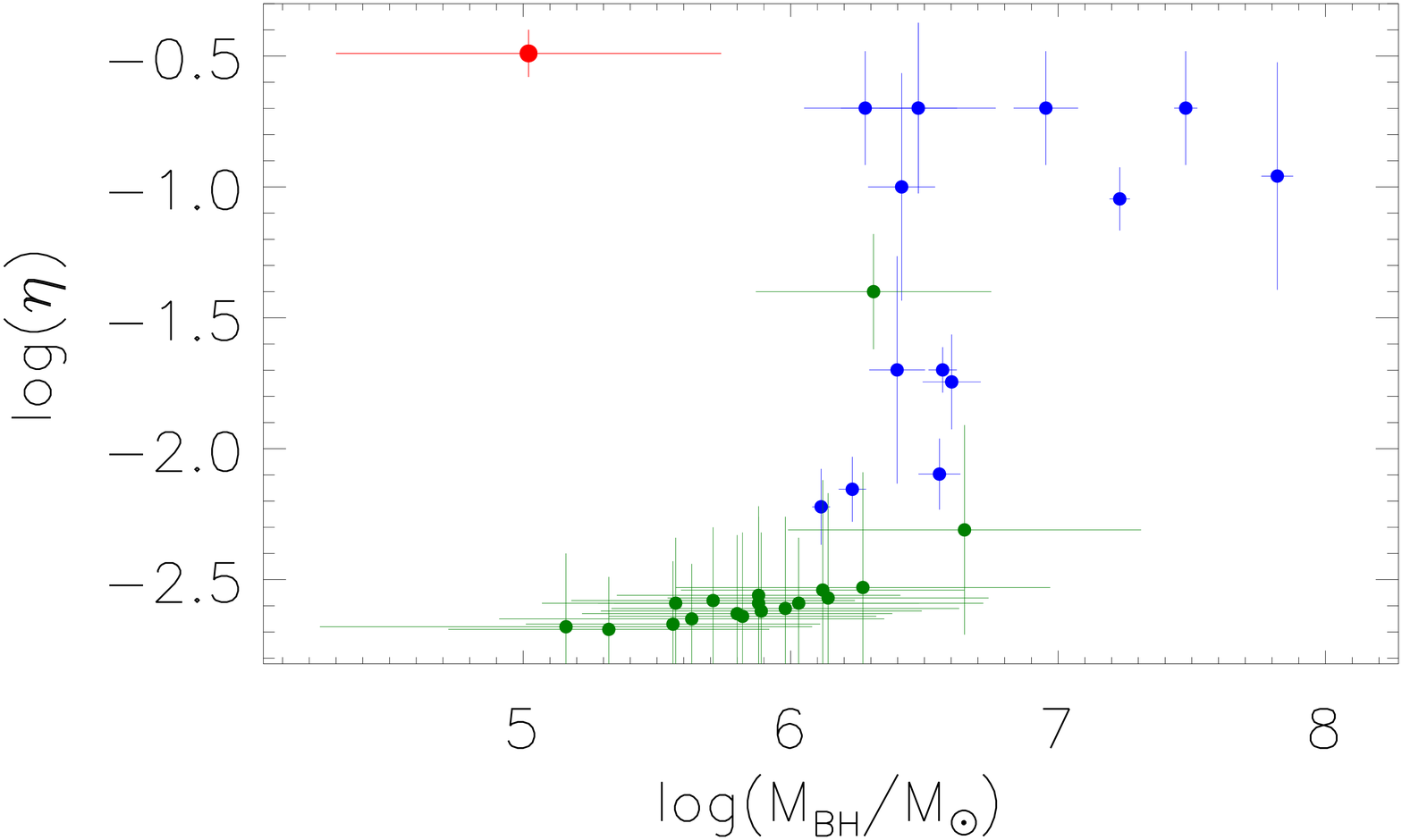}
\caption{On the dependence of energy transfer efficiency $\eta$ on central BH mass of the 
reported TDE candidates in \citet{mg19, zl21}. Symbols in blue represent the results collected from 
Table~2\ in \citet{mg19}, and symbols in dark green represent the results collected from Table~3\ in 
\citet{zl21}. The solid circle plus error bars in red show the results for the \obj~ determined in 
the manuscript.
}
\label{eff}
\end{figure}

	The well MCMC determined posterior distributions of the model parameters are shown in Fig.~\ref{par}, 
and accepted that $\log(M_{BH}/10^6M_\odot) = -0.98\pm0.14$ ($M_{BH}\sim1.05_{-0.29}^{+0.39}\times10^5{\rm M_\odot}$), 
$\log(M_*/M_\odot) = -1.89\pm0.05$, $\log(\beta) = -0.11\pm0.03$, $\log(\eta) = -0.49\pm0.09$, 
$\log(T_{vis}) = -1.29\pm0.13$, $\log(\Gamma) = 1.51\pm0.11$, $\log(k_{bol}) = 0.51\pm0.21$ and 
$n_\gamma=2.01\pm0.15$. The information of the model parameters are also listed in Table~1. 
The large value of $\eta\sim32\%$ can be well expected due to the relativistic jets in \obj. And the 
small value of $k_{bol}\sim3.2$ is also can be acceptable value, as the shown results in \citet{lc12} 
on ratios of bolometric luminosities to X-ray band luminosities of X-ray selected AGN. And the determined 
$\Gamma\sim32$ is a bit larger than the expected values around 10-20\ in \citet{bur11, ck12}, however, 
$\Gamma\sim32$ is one common value as the discussed results in \citet{cl07}. And, the determined value 
of $n_\gamma\sim2.01$ is well agreement with the expected global mean value as discussed in \citet{cl07}. 
All the determined model parameters have common values, and the BH mass is estimated as 
$M_{BH}\sim1.05\times10^5{\rm M_\odot}$ in \obj~ through the theoretical TDE model. The reported 
information of BH mass of \obj~ are shown in Fig.~\ref{all}, the determined BH mass 
$M_{BH}\sim1.05\times10^5{\rm M_\odot}$ in \obj~ is close to the previous results well discussed in 
\citet{pc15} but far from the other reports on BH mass discussed in \citet{ck12, stv17}.

\begin{table*}
\caption{Basic information of the model parameters applied to fit X-ray light curve of \obj}
\begin{center}
\begin{tabular}{ccccccccc}
\hline\hline
	& $\log(M_{BH}/10^6M_\odot)$  & $\log(M_*/M_\odot)$  &  $\log(\beta)$  & $\log(\eta)$  &  
	$\log(T_{vis})$ & $\log(\Gamma)$ & $\log(k_{bol})$ & $n_\gamma$ \\
\hline
\multicolumn{8}{c}{theoretical TDE model with $\gamma=4/3$ applied to describe the light curve 
	shown in left panel of Fig.~\ref{tde}} \\
\hline
limits	& [-3, 3] & [-2, 1.78] & [-0.22, 0.6] & [-2.3, -0.4] & [-3, 0] & [0, 1.7] & 
	[0.30, 2] & [1, 5] \\
$P_0$	& 0 & 0 & 0 & -0.8 & -1 & 1 & 1 & 2 \\
$P_M$ & $-0.98\pm0.14$ & $-1.89\pm0.05$ & $-0.11\pm0.03$ & $-0.49\pm0.09$ & $-1.29\pm0.13$ & 
	$1.51\pm0.11$ & $0.51\pm0.21$ & $2.01\pm0.15$ \\
\hline
\multicolumn{8}{c}{theoretical TDE model with $\gamma=4/3$ applied to describe the light curve 
	shown in middle panel of Fig.~\ref{tde}} \\
\hline
limits  & [-3, 3] & [-2, 1.78] & [-0.22, 0.6] & [-2.3, -0.4] & [-3, 0] & [0, 1.7] &
        [0.30, 2] & [1, 5] \\
$P_0$   & 0 & 0 & 0 & -0.8 & -1 & 1 & 1 & 2 \\
$P_M$ &	$-1.07\pm0.21$ & $-1.91\pm0.07$ & $-0.19\pm0.01$ & $-0.40\pm0.07$ & $-1.49\pm0.18$ &
	        $1.50\pm0.09$ & $0.42\pm0.26$ & $2.86\pm0.12$ \\
\hline
\multicolumn{8}{c}{theoretical TDE model with $\gamma=5/3$ applied to describe the light curve 
	shown in right panel of Fig.~\ref{tde}} \\
\hline
limits  & [-3, 3] & [-2, 1.78] & [-0.3, 0.4] & [-2.3, -0.4] & [-3, 0] & [0, 1.7] &
        [0.30, 2] & [1, 5] \\
$P_0$   & 0 & 0 & 0 & -0.8 & -1 & 1 & 1 & 2 \\
$P_M$ & $-1.59\pm0.33$ & $-1.85\pm0.09$ & $-0.223\pm0.02$ & $-1.27\pm0.21$ & $-1.18\pm0.11$ &
		        $1.44\pm0.14$ & $1.17\pm0.36$ & $2.62\pm0.19$ \\
\hline
\end{tabular}\\
\end{center}
Notice: There are nine columns in the table. The second to the last columns show the information of the 
eight model parameters in the theoretical TDE model discussed in Section 2. The row with the fist column 
named "limits" shows the information of the accepted limits of the prior uniform distributions of the 
model parameters. The row with the first column named "$P_0$" shows the starting values of the model 
parameters, when the procedure is running to fit the X-ray light curve. The row with the first column 
named "$P_M$" shows the determined values of the model parameters through the MCMC determined posterior 
distributions. \\
The third to the fifth rows show the basic information of the model parameters in the theoretical TDE 
model with $\gamma=4/3$ applied to describe the light curve shown in left panel of Fig.~\ref{tde}, with 
the corresponding MCMC determined posterior distributions of the model parameters shown in Fig.~\ref{par}. \\
The seventh to the ninth rows show the basic information of the model parameters in the theoretical TDE 
model with $\gamma=4/3$ applied to describe the light curve including the additional data point with 
JD=2456049.048 shown in middle panel of Fig.~\ref{tde}, with the corresponding MCMC determined posterior 
distributions of the model parameters shown in Fig.~\ref{p3ar}. \\
The eleventh to the thirteenth rows show the basic information of the model parameters in the theoretical 
TDE model with $\gamma=5/3$ applied to describe the light curve shown in right panel of Fig.~\ref{tde}, 
with the corresponding MCMC determined posterior distributions of the model parameters shown in 
Fig.~\ref{p2ar}. \\
\end{table*}

	Moreover, three points are noted, before giving further discussions on the best-fitting results 
and the determined model parameters by the theoretical TDE model.

	First and foremost, the circularization processes in TDEs are not considered in \obj. More recent 
discussions on theoretical circularization processes can be found in \citet{hs16, zo20, lo21}. Observational 
circularization emissions in TDEs have been reported and discussed in the unique TDE candidate ASASSN-15lh 
in \citet{lf16}, due to the two clear peaks detected in the UV band light curves. However, as the shown 
light curve in Fig.~\ref{tde} of \obj, there are no re-brightened peaks. Certainly, \citet{do16} have 
previously reported the ASASSN-15lh as a highly super-luminous supernova. If the ASASSN-15lh is not a TDE, 
the second UV peak in ASASSN-15lh is therefore not an unambiguous signature of delayed circularization. 
However, as commonly discussed in \citet{mg19, ji16, lo21, zl21}, common circularization emissions can 
lengthen the rise-to-peak timescale in TDEs expected variabilities. The lengthened rise-to-peak timescale 
can lead to apparently large $T_{vis}$, when the theoretical TDE model is applied to describe the variabilities. 
However, the determined parameter $\log(T_{vis})=-1.29\pm0.13$ can lead the expected viscous 
timescale $T_{vis}$ in \obj~ to be about $0.05days$, after considering the scaling factor 
$M_{\rm BH,~6}^{0.5}M_{\star}^{-1}R_{\star}^{1.5}$ shown in equation (6). The small value of $0.05days$ 
strongly indicates few effects of circularization emissions on our final results in \obj. Therefore, the 
simple case is mainly considered in \obj~ that the fallback timescales of the circularization processes 
are significantly smaller than the viscous timescales of the accretion processes, and the fallback materials 
will circularize into a disk as soon as possible.

	Besides, the light curve with data points shown as solid circles in left panel of Fig.~\ref{tde} 
has time durations about 179days in observer frame, however, there is an additional data point with 
JD=2456049.048\ and X-ray flux $0.17\times10^{-12}{\rm erg/s/cm^2}$ in \citet{pc15} not included in the 
light curve. Middle panel of Fig.~\ref{tde} shows the light curve including the additional data point 
shown as solid blue triangle. Based on the following two points, there are no further considerations and 
discussions on the additional data point with JD=2456049.048. On the one hand, the time dependent X-ray 
flux variabilities {\bf at late times} should be roughly described by $t^{-5/3}$, however, the time dependent 
X-ray flux variabilities from JD=2455885 to JD=2456049 (time duration about 164days) are simply described 
by $t^{\sim0}$. On the other hand, if the additional data point with JD=2456049.048 is included in the light 
curve to be described by the theoretical TDE model, best-fitting results and corresponding 99\% confidence 
bands shown in middle panel of Fig.~\ref{tde} can also be determined based on the same prior distributions 
of the model parameters, however, the fitting results are worse for the data points with JD smaller than 
2455887. And the posterior distributions of the model parameters are shown in Fig.~\ref{p3ar} with determined 
central BH mass about $\log(M_{BH}/10^6M_\odot)\sim-1.07\pm0.21$, which is similar as the determined BH mass 
$\log(M_{BH}/10^6M_\odot) = -0.98\pm0.14$ without consideration of the additional data point with 
JD=2456049.048. Here, there are no further discussions on the model parameters shown in Fig.~\ref{p3ar} 
and listed in Table~1. The main objective to show the posterior distributions of the model parameters is 
to assure acceptable model parameters through the theoretical TDE model applied to describe the X-ray 
light curve including the additional data point with JD=2456049.048. Therefore, the additional data point 
with JD=2456049.048 is not considered and discussed any more in the manuscript.

	Last but not the least, besides the best fitting results to the long-term X-ray variabilities 
in left panel of Fig.~\ref{tde} determined by the theoretical TDE model with the polytropic index 
$\gamma=4/3$, similar best fitting results to the X-ray light curve can also be determined by application 
of theoretical TDE model with the polytropic index $\gamma=5/3$. The same prior uniform distributions 
and starting values are accepted to the model parameters, besides a prior uniform distribution of the 
impact parameter of $\log(\beta)$ from $\log(0.50)$ to $\log(2.5)$ for main sequence stars with polytropic index of $\gamma=5/3$. 
The determined best fitting results and corresponding 99\% confidence bands are shown in right panel of 
Fig.~\ref{tde} to the X-ray light curve, and the posterior distributions of the model parameters are 
shown in Fig.~\ref{p2ar} and the accepted values listed in Table~1. The determined central BH mass is about 
$\log(M_{BH}/10^6M_\odot)\sim-1.59\pm0.33$, a bit smaller BH mass determined by the model with $\gamma=4/3$. 
However, considering the following main reason, the results determined by the model with $\gamma=5/3$ are 
not preferred. The determined energy transfer efficiency $\eta$ is about $\eta~\sim~0.054_{-0.021}^{+0.033}$, 
indicating a central non-spinning Schwarzschild BH, not reasonable to explain the relativistic jet in 
\obj. Therefore, in the manuscript, rather than the TDE model with $\gamma=5/3$, the theoretical TDE 
model with $\gamma=4/3$ is preferred to determine the final best descriptions to the long-term X-ray 
variabilities in \obj.

\subsection{Further discussions}

	First and foremost, based on the reported results in \citet{ck12} and in \citet{stv17}, previously 
expected BH mass could be larger than $2\times10^{7}{\rm M_\odot}$ and smaller than $10^8{\rm M_\odot}$. 
Here, we can check whether so large BH mass could be reasonable in \obj. Unfortunately, once accepted 
the prior uniform distribution of BH mass $\log(M_{\rm BH}/10^6{\rm M_\odot})$ from $\log(20)$ to $\log(100)$, 
the observed X-ray light curve of \obj~ can not be described by the model parameters within the accepted 
prior distributions. Therefore, the central BH mass of \obj~ is preferred to be about 
$1.05_{-0.29}^{+0.39}\times10^5{\rm M_\odot}$, through the theoretical TDE model. The determined BH mass 
well consistent with the results discussed in \citet{pc15}.

	Besides, we do not consider properties of the previously reported steep drop (or turnoff) in 
the X-ray variabilities at late times in \obj, such as discussed results in \citet{pc15}, which would 
be related to the transition from super-Eddington to sub-Eddington accretion as discussed in \citet{pc15} 
or to accretion disk transitioned from a thick disk to a thin disk as suggested in \citet{mb16} for the 
{\it Swift} J1644+57. Therefore, the proposed TDE model above cannot be applied to explain or to anticipate 
the steep drop at late times in \obj, but the collected long-term X-ray variabilities not including the 
steep drop features can provide sufficient information to determine the central BH mass by the theoretical 
TDE model.

	Last but not the least, it is a better idea to apply the same TDE model parameters to describe 
the long-term optical light curve of \obj, in order to confirm the reliability of the determined TDE 
model parameters, especially the central BH mass. Fortunately, there are 7 reliable data points in 
each optical $ugriz$ bands in \citet{pc15}. Then, we check whether the TDE  model expected best 
descriptions to the X-ray light curve can be applied to describe the optical light curve. Here, the 
g-band light curve in observer frame is collected from Table~2\ in \citet{pc15} and 
shown in Fig.~\ref{opt}. However, there are quite few data points in the optical g-band light curve 
of \obj, and there is not clear peak information of the optical light curve, not similar as the X-ray 
light curve of \obj. Therefore, not the detailed MOSFIT code or the TDEFIT code is applied to described 
the optical light curve of \obj, but the TDE model determined best descriptions to the X-ray light 
curve weakened by a factor about 1430 and shifted by -117days can be well applied to describe the 
optical light curve of \obj. The weakened factor of 1430 could be a reasonable value, such as the 
ratios round 1000 of X-ray luminosities to optical luminosities of \obj~ shown in Figure~1\ in 
\citet{pc15}. And the descriptions to the optical g-band light curve is shown in Fig.~\ref{opt}, 
indicating that the similar TDE model can be also applied to describe optical variabilities, indicating 
the determined TDE model parameters are reliable enough.

	Before the end of the section, further discussions are given on the determined central BH mass 
of \obj. Among the reported central BH masses of TDE candidates in \citet{gm14, mg19}, the \obj~ has 
the smallest central BH mass. Even considering the more recent reported BH masses  of TDE candidates 
in \citet{zl21} which are systematically smaller than the BH masses estimated in \citet{mg19}, the 
\obj~ also has the smallest central BH mass, accepted mean BH mass $1.15\times10^5{\rm M_\odot}$ in 
iPTF16fnl in \citet{zl21}. Meanwhile, among the reported energy transfer efficiencies $\eta$ of TDE 
candidates in \citet{gm14, mg19, zl21}, the \obj~ has the largest energy transfer efficiency $\eta$. 
Fig.~\ref{eff} shows the dependence of energy transfer efficiency $\eta$ on central BH mass of the 
reported TDE candidates in \citet{mg19, zl21}. Here, only the main values of iPTF16fnl and GALEX D23H-1 
are collected from the Table~3\ in \citet{zl21}. The results shown in Fig.~\ref{eff} strongly indicate 
that the \obj~ is one unique TDE candidate among the reported TDE candidates. Furthermore, the 
inconsistent dependence of energy transfer efficiency on central BH mass between the \obj~ and the other 
TDE candidates provides further clues to detect and/or anticipate candidates of relativistic TDE to 
make the birth of relativistic jets.

\section{Main Summaries and Conclusions}

   Finally, our main summaries and conclusions are as follows.
\begin{itemize}   
\item Long-term X-ray variabilities of \obj~ with JD from 2455708.915 to 2455887.787 can be well described 
	by a main-sequence star with $\gamma=4/3$ tidally disrupted by a central BH with mass about 
	$\log(M_{BH}/10^6M_\odot) = -0.98\pm0.14$ and with energy transfer efficiency about 32\%, to 
	support a central rapid spinning low mass BH in \obj.
\item Long-term X-ray variabilities of \obj~ with JD from 2455708.915 to 2455887.787 can also be well described 
	by a main-sequence star with $\gamma=5/3$ tidally disrupted by a central BH with energy transfer 
	efficiency about 5\%, indicating a central non-spinning low mass BH. Considering the relativistic 
	jet in \obj, the determined model parameters with $\gamma=5/3$ are not preferred.
\item Considering the additional data point with JD=2456049.048, there are similar central BH masses in 
	\obj~ determined by applications of theoretical TDE model to describe the X-ray light curve 
	including the additional data point.
\item Based on the long-term X-ray variability properties well described by the theoretical TDE model in \obj, 
	the central BH mass can be well determined as $1.05_{-0.29}^{+0.39}\times10^5{\rm M_\odot}$, 
	after well considering the relativistic beaming effects. 
\item Based on the reported BH mass and energy transfer efficiency of TDE candidates, the \obj~ is an unique 
	TDE candidate in the space of BH mass versus energy transfer efficiency, providing further clues to 
	detect and/or anticipate candidates of relativistic TDEs to make the birth of relativistic jets.
\end{itemize}

\section*{Acknowledgements}
Zhang gratefully acknowledges the anonymous referee for giving us constructive 
comments and suggestions to greatly improve our paper. Zhang gratefully acknowledges the kind 
support of Starting Research Fund of Nanjing Normal University and the kind financial support 
from NSFC-12173020. The paper has made use of the public code of TDEFIT 
(\url{https://github.com/guillochon/tdefit}) and MOSFIT (\url{https://github.com/guillochon/mosfit}).



\end{document}